\documentclass[3p]{elsarticle}

\usepackage{hyperref}

\journal{Elsevier}
\usepackage{graphicx}
\usepackage{titlesec}
\usepackage{graphicx}
\usepackage[most]{tcolorbox} 
\usepackage[ansinew]{inputenc}
\usepackage{array}
\usepackage{ragged2e}
\usepackage{makecell}
\usepackage{amsthm}
\usepackage{amsfonts}
\usepackage{color}
\usepackage{soul}
\usepackage{tabularx}
\usepackage{adjustbox}
\usepackage{adjustbox,lipsum}
\usepackage{rotating}
\usepackage{xcolor,colortbl}
\usepackage{subfiles}
\usepackage{multicol}
\usepackage{stfloats}
\usepackage{url}
\definecolor{Gray}{gray}{0.9}
\usepackage{balance}
\usepackage{xcolor}
\usepackage{amsmath}
\usepackage{tikz}
\usetikzlibrary{shapes}
\newcolumntype{C}[1]{>{\centering\let\newline\\\arraybackslash\hspace{0pt}}m{#1}}
\usepackage{rotating}
\usepackage{pdflscape}
\usepackage{afterpage}
\usepackage{capt-of}
\usepackage{longtable}
\usepackage{stfloats}
\usepackage{amsmath,amssymb,amsfonts}
\usepackage{algorithmic}
\usepackage{graphicx}
\usepackage{caption}
\usepackage{textcomp}
\usepackage{longtable}
\usepackage{enumitem}
\usepackage{rotating}
\usepackage{subfigure}
\usepackage{hyperref}
\usepackage{amsxtra}
\usepackage{natbib}
\usepackage{amssymb}
\usepackage{subfigure}
\usepackage{latexsym}
\usepackage{caption}
\usepackage{amsmath,amsfonts,amssymb,amsthm}
\usepackage[ruled,vlined]{algorithm2e}
\usepackage{multirow}
\definecolor{Gray}{gray}{0.9}
\usepackage{balance}
\usepackage{etoolbox}

\makeatletter
\def\ps@pprintTitle{%
   \let\@oddhead\@empty
   \let\@evenhead\@empty
   \def\@oddfoot{\reset@font\hfil\thepage\hfil}
   \let\@evenfoot\@oddfoot
}
\makeatother

\begin{document}
\begin{frontmatter}

\title{Centralized Defense: Logging and Mitigation of Kubernetes Misconfigurations with Open Source Tools}

\author[1,2]{Eoghan Russell}
\ead{eoghanrussell1212@gmail.com}
\author[2]{Kapal Dev}
\ead{kapal.dev@ieee.org}

\address [1] {Intel, Ireland}
\address [2]{Department of Computer Science, Munster Technological University,Ireland}

\begin{abstract}
Kubernetes, an open-source platform for automating the deployment, scaling, and management of containerized applications, is widely used for its efficiency and scalability. However, its complexity and extensive configuration options often lead to security vulnerabilities if not managed properly. This paper presents a detailed analysis of misconfigurations in Kubernetes environments and their significant impact on system reliability and security. A centralized logging solution was developed to detect such misconfigurations, detailing the integration process with a Kubernetes cluster and the implementation of role-based access control. Utilizing a combination of open-source tools, the solution systematically identifies misconfigurations and aggregates diagnostic data into a central repository. The effectiveness of the solution was evaluated using specific metrics, such as the total cycle time for running the central logging solution against the individual open source tools.

\end{abstract}

\end{frontmatter}
\section{Introduction} 

Kubernetes, an open-source platform for automating the deployment, scaling, and management of containerized applications, is widely used for its efficiency and scalability. However, its complexity and extensive configuration options often lead to security vulnerabilities if not managed properly. Common issues in Kubernetes include excessive permissions, misconfigured network policies, and insecure default settings, all of which can create potential entry points for attackers. Ensuring robust security practices and continuous monitoring is essential to maintaining the integrity and reliability of Kubernetes clusters. Among these challenges, misconfigurations are a significant concern that can expose cloud-native applications to severe security threats.

While many open-source tools are available to identify misconfigurations within Kubernetes manifest files, a central logging solution does not exist. This paper proposes such a central logging solution which identifies misconfigurations and provides advisory defensive solutions regarding the misconfigurations identified. The results are then presented to a cluster administrator via the home page of the central logging solution.

The need for this central logging solution is motivated by the awareness that Kubernetes holds a pivotal position in the deployment of cloud-native applications and the serious risks posed by misconfigurations in Kubernetes manifest files. This awareness is informed by this author's daily usage of Kubernetes providing firsthand knowledge of the consequences of misconfigurations, the vulnerabilities they might introduce, to the avenues they open for potential system breaches by attackers.

This paper proposes to identify the vulnerabilities associated with common misconfigurations and their potential impact on the Kubernetes environment.  Additionally, it will determine the efficacy of the most popular open-source tools which specialise in the detection of Kubernetes misconfigurations.  The proposed central logging solution will be evaluated for how effectively it aids in the mitigation of common security misconfigurations and its ability to seamlessly integrate with popular open-source tools. The delta in latency compared to running the individual open-source tools will be determined against viewing the results from the centralised logging solution.

This paper delves into Kubernetes security, focusing on how misconfigurations can create vulnerabilities within a Kubernetes environment. Section 1 introduces the paper, highlighting the security risks of Kubernetes misconfigurations, the lack of a central logging solution, and the research objectives this paper aims to address. Section 2 explores security vulnerabilities within Kubernetes, including real-world attack scenarios. It examines how Kubernetes misconfigurations can lead to significant issues and their link to security breaches, concluding with a review of open-source tools for detecting these misconfigurations. Section 3 explains the architecture of the proposed central logging system. Section 4 details the implementation of the central logging system using the selected tools. It evaluates the logging solution by analyzing metrics such as execution time from start to finish, frequency of reruns, and effectiveness of comparing outputs from different tools. Additionally, it offers a comparative analysis of the tools based on personal insights, highlighting their benefits and drawbacks. Finally, Section 5 concludes the paper by reviewing the implemented solution, discussing potential areas for further improvement, and providing a final observation.

The primary contributions of this paper are the development of a centralized logging solution to detect Kubernetes misconfigurations, the integration of multiple open-source tools to enhance the detection capabilities, the evaluation of the solution's effectiveness in real-world scenarios, and a comparative analysis of various open-source tools for identifying Kubernetes misconfigurations.

\section{Literature Review}

Academic research has highlighted security concerns with Kubernetes, posing risks to its adoption despite its advantages as a lightweight, efficient, and cost-effective solution \cite{mondal2022kubernetes}. Enterprises have sought to identify and address these concerns after several high-profile attacks have occurred. According to a survey conducted by CNCF in 2022, lack of user training and security of the Kubernetes environment are the top challenges faced when deploying containers, 40\% of the 2063 respondents reported that security is their biggest challenge in production environments \cite{cncf_2023_survey}. A StackRox survey conducted in 2020 suggests that 44\% of organisations delay deployment due to security concerns \cite{stackrox_survey_2020}. The results also states that 94\% of organisations experienced a security incident within the previous 12 months \cite{stackrox_survey_2020}.  Two attacks which were highly reported by the news media occurred to Tesla and Capital One,  both exposed vulnerabilities in the system structure. In 2018, Tesla experienced an intrusion due to a lack of security best practices, the Kubernetes console was not password-protected and malicious users were able to gain AWS credentials that were exposed in a container \cite{goodin_tesla_2018}. At Capital One, an attack on a Kubernetes cluster was found to have been caused by the misconfiguration of the firewall, allowing an attacker to query internal metadata \cite{techgenix_security_incidents}. 

A paper proposed by Shamim et al. discusses the growing concern of security vulnerabilities within a Kubernetes environment, through a systematic analysis of internet artifacts that led to the identification and categorization of 11 essential security practices. Their research identified key measures such as robust authentication, authorization, and continuous updates as critical to enhancing the security posture of Kubernetes deployments. Moreover, their research highlights the importance of a comprehensive, informed approach to safeguard sensitive data and infrastructure against evolving cyber threats, providing a crucial roadmap for practitioners aiming to mitigate vulnerabilities within these complex containerised infrastructures \cite{shamim2020commandments}.

Yang et al. provides a crucial examination of security vulnerabilities in Kubernetes, pinpointing how excessive permissions granted to third-party applications can lead to critical vulnerabilities, enabling attackers to potentially seize control of entire Kubernetes clusters. Their investigation, covering all third-party applications listed by the Cloud Native Computing Foundation (CNCF) and services provided by leading cloud vendors, uncovers that a significant portion of these applications carry inherent risks due to overly broad permissions. The study's impact is highlighted by the identification of eight new Common Vulnerabilities and Exposures (CVEs), demonstrating the serious security implications and underscoring the need for rigorous permission controls within the Kubernetes ecosystem to mitigate potential attack vectors effectively \cite{yang2023take}.

\begin{figure}[h!]
	{\includegraphics[width=\textwidth]{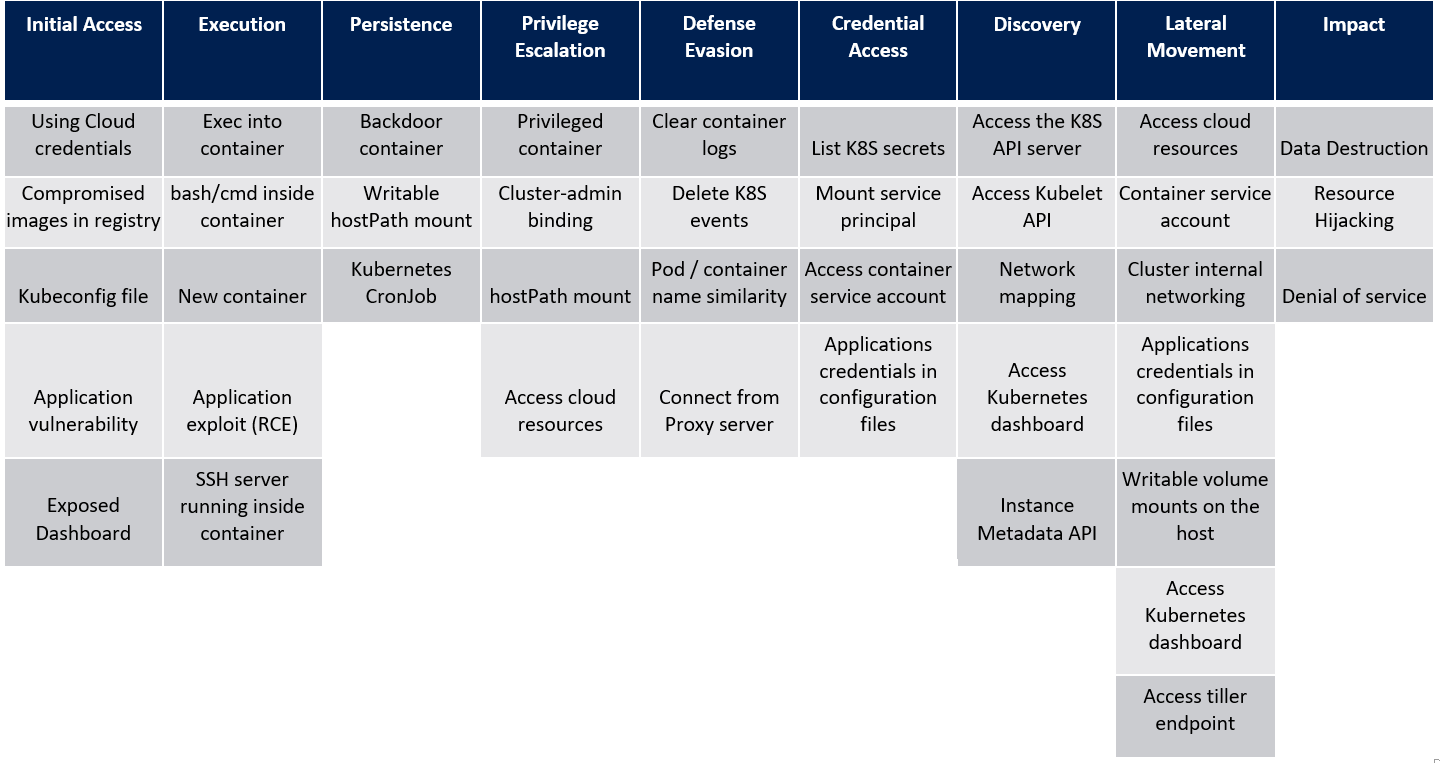}}
	\caption{The Kubernetes attack matrix enumerates various methods attackers employ to accomplish the nine tactics specified in the Mitre ATT\&CK framework \cite{noauthor_mitre_2024}: Initial Access, Execution, Persistence, Privilege Escalation, Defense Evasion, Credential Access, Discovery, Lateral Movement, and Impact. \cite{weizman_threat_2020}}
\end{figure}

The Kubernetes Threat Matrix, developed by the Azure Security Centre and based on the Mitre ATT\&CK framework \cite{noauthor_mitre_2024}, outlines the complex vulnerabilities of Kubernetes clusters \cite{weizman_threat_2020}. This matrix is a pivotal resource, offering a structured analysis of potential cybersecurity threats and the corresponding tactics attackers employ to infiltrate and compromise Kubernetes environments. Each tactic outlined within the matrix corresponds to a phase in the attack life cycle, from initial access to eventual impact, providing a comprehensive view of the attack vectors that adversaries might exploit. By leveraging the Mitre ATT\&CK framework, a well-regarded knowledge base of cyber attack techniques, the Kubernetes Threat Matrix offers invaluable insights into securing Kubernetes clusters against a wide array of cybersecurity threats.

\textbf{Initial Access:} This tactic involves an attacker obtaining access to the cluster through various means, such as exploiting legitimate credentials, exploiting vulnerabilities in exposed endpoints, or compromising container images and inducing their execution within the cluster.

\textbf{Execution:} Following access, the attacker executes malicious activities inside the cluster. This may involve the creation of a malicious container or executing commands within an existing container, using tools like \texttt{kubectl exec}, assuming the necessary permissions are obtained.

\textbf{Persistence:} This tactic ensures the attacker maintains access to the cluster even after initial entry points are closed. Techniques include deploying containers with embedded malicious code, known as backdoor containers, to facilitate re-entry.

\textbf{Privilege Escalation:} Attackers enhance their access levels to exploit more resources and sensitive information. This could involve accessing a container with elevated privileges or exploiting Role-Based Access Control (RBAC) to assume roles like cluster-admin, granting extensive control over cluster resources.

\textbf{Defense Evasion:} Attackers employ methods to conceal their presence, such as erasing logs or camouflaging malicious containers with conventional naming schemes to blend into the cluster environment.

\textbf{Credential Access:} This involves attackers targeting and stealing credentials stored within the cluster, like Kubernetes Secrets, to access sensitive information and further exploit the cluster.

\textbf{Discovery:} Attackers explore the cluster to understand its layout and identify targets. This may involve querying the API Server, Kubelet API, Kubernetes Dashboard, or using the internal network communications between pods.

\textbf{Lateral Movement:} With knowledge of the cluster's architecture, attackers move within the cluster to reach their objectives. Techniques for discovery can also facilitate movement, leveraging the cluster's internal communications or administrative interfaces.

\textbf{Impact:} The ultimate goal of the attacker is to disrupt the cluster's normal operations negatively. This could involve data deletion, configuration changes, using the cluster for cryptocurrency mining, or executing a denial of service (DoS) attack to deny legitimate users access to services.


\newcolumntype{C}{>{\centering\arraybackslash}X}
\begin{center}
\begin{table}[h!]
\begin{tabularx}{\textwidth}{|C|C|C|C|} 
 \hline
 \textbf{CVE} & \textbf{Cause} & \textbf{Effect} & \textbf{Consequences} \\
 \hline
 2014-0047 & Unprotected Resources & Exposed System & Gain Information \\ 
 \hline
 2014-3499 & Incorrect Permission Management & Exposed System & Gain Privileges \\
 \hline
 2014-5277 & Incorrect Recovery Mechanism & Restricted Violation & Gain Information \\
 \hline
 2014-5279 & Unprotected Resources & Exposed System & Gain Privileges \\
 \hline
 2014-5280 & Incorrect Permission Management & Restricted Violation & Execute code \\
 \hline
 2014-8179 & Incorrect Configuration & Exposed System & Execute code \\
 \hline
 2014-5282 & Improper Validation & Exposed System & Bypass \\
 \hline
 2014-6407 & Incorrect Permission Management & Restriction Violation & Execute code \\
 \hline
 2023-30512 & cfs-csi-cluster-role & Privilege Escalation & Access to entire cluster \\
 \hline
 2020-8559 & Unvalidated redirect on proxied upgrade requests & Privilege Escalation & Full cluster compromise \\
 \hline
\end{tabularx}
\caption{Common vulnerabilities, causes, effects, and consequences \cite{RN482}}
\end{table}
\end{center}

Table 1 presents an overview of common vulnerabilities, with a focus on the implications of misconfigurations within Kubernetes environments. The table categorizes these vulnerabilities by their Common Vulnerabilities and Exposures (CVE) identifiers, highlighting the causes, such as unprotected resources, incorrect permission management, and improper validation, which are prevalent in Kubernetes configurations. The effects and consequences range from exposed systems and restriction violations to full cluster compromise, emphasizing the critical impact misconfigurations can have on the security and integrity of Kubernetes clusters. Notably, vulnerabilities like 'cfs-csi-cluster-role' and 'Unvalidated redirect on proxied upgrade requests' directly illustrate how specific misconfigurations in Kubernetes can lead to severe consequences, including privilege escalation and complete cluster compromise.

Echoing the concerns highlighted in the table, the Red Hat 2023 Report reinforces the notion that misconfiguration remains a pressing challenge within Kubernetes environments, with 28\% of users citing misconfigurations and vulnerabilities as their top concerns \cite{noauthor_state_2023}. This apprehension is largely attributed to Kubernetes' extensive customization capabilities, which, while beneficial, introduce significant risk factors. Human error is also behind the vast majority of security breaches, which leads to the misconfigurations of production environments \cite{noauthor_dbir_nodate}.  

In the pursuit of fortifying Kubernetes deployments against security vulnerabilities, academic endeavors have meticulously explored the avenues through which system misconfigurations could be preemptively identified and rectified. Rahman et al.'s seminal work \cite{rahman2023security} stands at the forefront of this exploration, offering an empirical study that delves into the intricacies of security misconfigurations within Kubernetes manifests. By scrutinizing 2,039 Kubernetes manifest files sourced from 92 open-source repositories, they meticulously categorized various forms of security misconfigurations, highlighting the paramount importance of addressing these vulnerabilities to safeguard against potential security threats. Their research not only shed light on the prevalence of such misconfigurations but also led to the development of SLI-KUBE, a pioneering static analysis tool designed to quantitatively assess the frequency of these misconfigurations \cite{rahman2023security}.

This paper builds upon Rahman et al.'s foundational work, aiming to extend the discourse beyond the identification of misconfigurations by proposing a centralized logging solution. Such a solution is envisioned to seamlessly integrate with the Kubernetes environment, offering a comprehensive mechanism to alert system administrators of potential vulnerabilities across various layers of the Kubernetes stack. At the core of Kubernetes, pods represent the most fundamental unit, encapsulating one or more containers that collectively embody an application instance \cite{nguyen2020horizontal}. Configured through YAML-formatted Kubernetes manifest files, these pods are susceptible to security misconfigurations that could compromise the integrity and security of the entire system \cite{rahman2023security}.

Building on the insights provided by Rahman et al. regarding the crucial role of static analysis tools in highlighting security misconfigurations within Kubernetes environments \cite{rahman2023security}, this paper proposes an advancement through the implementation of a centralized logging system. This proposed system seeks to offer a comprehensive perspective on the security health of Kubernetes deployments, empowering system administrators with the capabilities to rapidly detect and address potential security vulnerabilities. By amalgamating the detailed detection and classification of security misconfigurations, as outlined by Rahman et al., with a novel logging mechanism, this paper aims to bolster the security robustness of Kubernetes infrastructures in the face of continually emerging cybersecurity challenges. 

Following the comprehensive framework proposed by Rahman et al. \cite{rahman2023security}, Mahajan's model introduces a proactive strategy by enforcing real-time misconfiguration checks prior to deployment, ensuring that any detected issues must be rectified before proceeding \cite{mahajan2022detection}. While this model emphasizes preventative measures, it does not offer a mechanism for the centralized collection, logging, or storage of the findings from static analysis tools, highlighting a critical area for enhancement. The landscape of existing open-source tools, including Checkov, KubeLinter, Datree, Trivy, and Snyk \cite{RN480}\cite{RN481}, presents a robust arsenal for the identification and notification of prevalent misconfigurations. Yet, the necessity to operate these tools in isolation introduces significant challenges in terms of resource consumption, operational complexity, and the risk of security oversights due to potential omissions. This scenario underscores the imperative for a unified solution that amalgamates the insights garnered from these disparate tools into a cohesive, centralized logging system. Such a system is envisaged to streamline the process of managing misconfiguration vulnerabilities by serving as a comprehensive repository and analysis point for security data across the Kubernetes deployment. By addressing the fragmentation inherent in the current use of multiple tools, which may lead to redundant or conflicting recommendations, the central logging solution proposed in this paper aims to simplify the resolution process by synthesizing outputs and guiding system administrators towards effective remediation strategies. This integration seeks to close the gap left by the piecemeal approach of existing tools, providing a singular, efficient platform for enhancing the security posture of Kubernetes environments \cite{RN480} \cite{shamim2021mitigating} \cite{Bagheri2023} \cite{KermabonBobinnec2022}.

Beyond the initial identification of common security misconfigurations, further analysis into the types of vulnerabilities and the potential impacts on Kubernetes environments is essential. Misconfigurations such as the activation of hostIPC or inappropriate security contexts can lead to severe security implications, including cross-pod access and privilege escalation. These issues are further compounded when considering the dynamic and distributed nature of containerised environments, where the scale and speed of deployment can often outpace traditional security measures. Bose et al. emphasize that security-related defects in Kubernetes manifests are under-reported, suggesting a significant gap in the awareness and remediation of such vulnerabilities within the Kubernetes community \cite{bose2021under}. Their research underscores the critical need for comprehensive security practices and the development of tools to systematically identify and mitigate security defects in Kubernetes configurations.

The risks associated with misconfigured Kubernetes manifests are not purely theoretical. There have been documented instances where such vulnerabilities have been exploited to conduct attacks ranging from data theft to the disruption of service. The work by Shamim et al. explores how malicious actors can leverage misconfigurations, such as those related to security best practices violations, to perform attacks like denial of service against Kubernetes clusters \cite{shamim2021mitigating}. This research sheds light on the practical applications of theoretical vulnerabilities, reinforcing the importance of strict configuration management and the implementation of security best practices to safeguard against potential exploits.

The intricate nature of Kubernetes environments significantly amplifies the difficulties in ensuring secure configurations are maintained. Navigating through the myriad of options within Kubernetes manifest files demands a thorough comprehension of their potential security impacts. To uphold the security integrity of Kubernetes clusters, it's imperative to employ tools and methods that assist in identifying and correcting misconfigurations. A host of tools, including Kubesec, Kube-hunter, and Trivy, have been developed specifically to tackle cluster misconfigurations, playing a pivotal role in bolstering Kubernetes security \cite{karakacs2023enhancing}.

However, the concept of integrating a unified logging framework, one that can consolidate outputs from a variety of such tools into a single, unified system, remains largely uncharted. This gap becomes even more pronounced when considering the challenges associated with presenting logged information in Kubernetes in a way that is both efficient and user-friendly for developers \cite{koryugin2023analysing}. Although research efforts, like the one by Horalek et al. \cite{horalek2022proposed}, have ventured into developing systems for aggregating and analyzing application logs within Kubernetes, these endeavors have mainly focused on operational aspects, overlooking the critical angle of security. Horalek et al.'s study, despite recognizing the complexity involved in delivering comprehensive logging solutions for Kubernetes applications, primarily highlighted operational benefits rather than delving into security ramifications. This oversight accentuates the urgent necessity for a centralized security logging system, aimed at merging and deciphering security-related data from various tools. Such a system would provide a comprehensive overview of the security posture within Kubernetes environments, addressing a crucial void in current research.

The open-source community has developed a plethora of tools specifically designed to identify misconfigurations in Kubernetes manifest files. These tools offer comprehensive scanning capabilities, enabling users to detect vulnerabilities, adherence to best practices, and potential security loopholes within their Kubernetes configurations. 

\textbf{Trivy} is a comprehensive container and Kubernetes vulnerability scanner. It quickly scans container images and Kubernetes manifests to identify vulnerabilities, misconfigurations, and outdated dependencies using an extensive vulnerability database. Trivy integrates smoothly with Kubernetes, facilitating automated security checks across development and deployment workflows. It's a critical tool for maintaining the security of Kubernetes environments \cite{trivy}.

\textbf{Kubesec} evaluates Kubernetes manifests to ensure they meet security best practices. It checks configurations against a series of security benchmarks, focusing on pod security policies, the risk of privilege escalation, and network policies. Kubesec offers actionable insights and remediation advice, helping users secure their Kubernetes deployments more effectively \cite{kubesec}.

\textbf{Kube-score} is a tool that analyzes Kubernetes manifests against best practices and standards. It provides a score and detailed feedback on configuration quality and reliability, identifying potential improvements and misconfigurations. kube-score guides developers and administrators in refining their Kubernetes configurations for better adherence to industry standards \cite{kube-score}. 

\textbf{Kubeaudit} is a Kubernetes-specific auditing tool that conducts thorough checks for security, reliability, and compliance. It uses a wide range of audit rules to examine Kubernetes manifests for misconfigurations and compliance issues, also supporting customizable rules for organization-specific audits. Kubeaudit is essential for ongoing auditing and compliance assurance in Kubernetes environments \cite{kubeaudit}. 

\textbf{KubeLinter} is a static analysis tool designed to identify misconfigurations in Kubernetes manifests and Helm charts. It checks for common configuration errors and flags the potential issues. 
\newpage
Each tool has their own interpretation of how severe a misconfiguration is to be treated, this is how these 4 tools treated severity levels in terms of misconfigurations:
\begin{itemize}
	\item \textbf{Trivy:} Trivy classified its severity levels as Critical, High, Medium, Low.
    \item \textbf{Kube-score:} Kube-score evaluates configuration issues using a grading scale. These grades were converted into severity levels based off their grade where a grade of 10 is labeled as "CRITICAL," grades between 7 and 9.99 are considered "HIGH," grades between 5 and 6.99 are "MEDIUM," and grades below 5 are "LOW." Grades not fitting these categories are marked as "UNKNOWN."
    \item \textbf{Kube-linter:} Kube-linter does not assign explicit severity levels in its JSON output, focusing instead on the presence and specifics of misconfigurations.
    \item \textbf{Kubesec:} Kubesec assigns scores that can indirectly imply severity; higher scores indicate better security practices, whereas lower scores suggest critical vulnerabilities needing immediate attention.
    \item \textbf{Kubesec:} Kubesec assigns severity based on a point system, these points were converted into severity categories where a score of -30 is considered "CRITICAL," scores of -9 and -7 are "HIGH," scores of -3 and -1 are "MEDIUM," and scores of 1 and 3 are "LOW." Scores not explicitly listed are categorized as "UNKNOWN."
\end{itemize}

\section{Architectural Design}
This section explains the design of the proposed central logging solution and the proposed flow of the central logging solution. 

\begin{figure}[h!]
	\centering
	{\includegraphics[width=\textwidth]{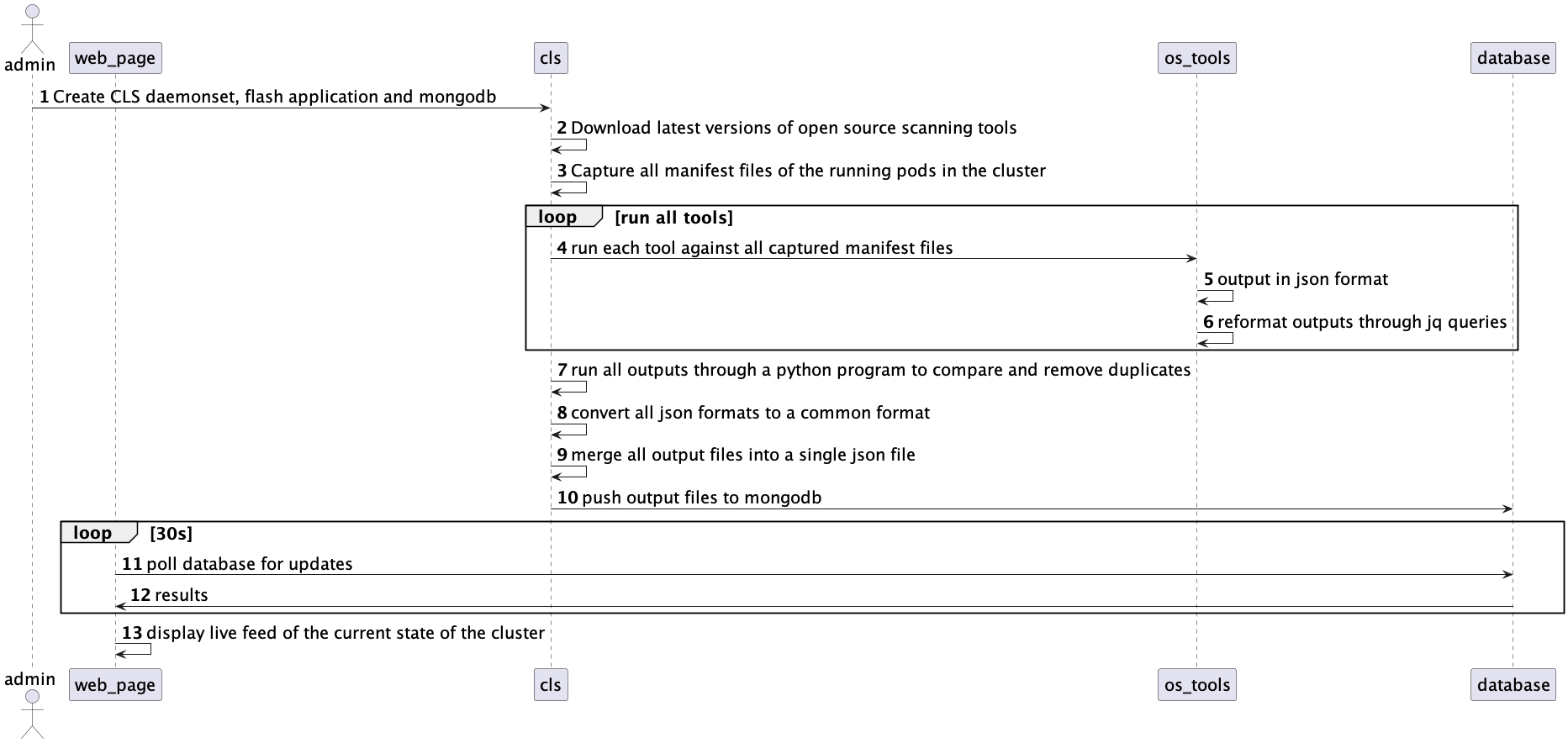}}
	\caption{Central logging solution flow }
\end{figure}

\newpage
Figure 2 depicts the proposed flow for the central logging solution. 
The flow is as follows:
\begin{enumerate}
\item  An administrator will deploy the central logging solution, flask application and database in a Kubernetes environment. 
\item The cls  downloads the latest versions of the open source scanning tools.
\item The cls captures all manifest files of the running pods in the cluster and stores these in json files. 
\item In a loop, each tool is ran against each manifest file.
\item The output is in json format. 
\item The cls then reformats all json files through jq to only capture the required fields. 
\item The cls passes all json output files to a comparative python program to try and remove duplicates across the tools. 
\item All outputs are then converted into a common template. 
\item All outputs are then combined into a single json file, representing 1 per live pod in the cluster.
\item The cls  pushes all resulting files to the mongodb. 
\item  In a loop, the flash application is constantly polling the database for updates, refreshing every 30 seconds. 
\item The web page displays a live feed of the current state of the cluster, and all misconfigurations and issues found across all the tools. 
\end{enumerate}

\section{Implementation}

The central logging solution is composed of several Docker containers: one for running the tools and another for the Flask application managing the web interface. The Central Logging Solution Dockerfile utilizes \texttt{python:3.12.2-alpine3.18} as a lightweight environment, setting \texttt{/app} as the workspace. It copies necessary bash scripts (\texttt{download\_tools.sh}, \texttt{run\_script.sh}, \texttt{entrypoint.sh}, \texttt{convert\_jq.sh}, \texttt{merged.sh}) and Python files (\texttt{compare.py}, \texttt{sender.py}) to the container, grants execution permissions to the copied scripts, installs compilers (\texttt{g++}, \texttt{gcc}), libraries (\texttt{libffi-dev}, \texttt{musl-dev}, \texttt{openblas-dev}, \texttt{lapack-dev}), and tools (\texttt{bash}, \texttt{curl}, \texttt{jq}), installs \texttt{kubectl} for cluster communication, installs \texttt{pymongo} for database interaction, and sets \texttt{/app/entrypoint.sh} as the entry point script.

The Flask Application Dockerfile uses \texttt{python:3.9} as the base image, sets \texttt{/app} as the workspace, copies the Flask application code, installs Flask and \texttt{Flask-PyMongo}, exposes port 5002 for the Flask application, sets environment variables for Flask execution, and runs the Flask application using \texttt{flask run}.

The Kubernetes cluster was deployed using Docker Desktop, integrating Docker's containerization with Kubernetes' orchestration capabilities. Role-Based Access Control (RBAC) is essential for secure access to the cluster's resources. The setup includes Trivy Cluster Role Binding, which grants permissions to the default service account in the \texttt{security-namespace}, and Custom Cluster Role Binding, which links the view cluster role with a custom service account in the \texttt{security-namespace}.

The entrypoint script executes the following sequence: downloads and installs the latest versions of the tools, captures manifest files of all pods in the cluster, runs each tool against the manifests and formats the results, compares the results to remove duplicates, converts the outputs to a common template, merges all outputs into a single JSON file per pod, and uploads the JSON files to MongoDB.

The \texttt{run\_script.sh} script captures manifest files and executes each tool. It initializes directories for outputs, retrieves namespaces and iterates through each to list and process all pods, runs tools (\texttt{Kubesec}, \texttt{Trivy}, \texttt{Kube-score}, \texttt{Kube-linter}) against each pod manifest and stores results in JSON format, and processes outputs using \texttt{jq} for formatting and storage.

The \texttt{compare.py} script eliminates duplicate results. It sets up directories and fetches file lists, loads common files and analyzes similarities using Python's \texttt{SequenceMatcher}, and stores unique issues in lists for further processing.

A Bash script uses \texttt{jq} to convert JSON files to a uniform structure and merges them into a single file per pod. A Python script automates importing JSON files into MongoDB, ensuring no duplicate data is uploaded.

The Flask application interfaces with MongoDB to present information about security vulnerabilities. The main index page displays an overview of vulnerabilities by severity, the severity level page lists vulnerabilities by specific severity levels, and the collection page provides detailed views of vulnerabilities in each collection. The application showcases various web pages: the CLS homepage displays the cluster's current status and vulnerability overview, the severity level page shows detailed information about high-severity vulnerabilities, and the collection page details security misconfigurations for individual collections.

\subsection{Showcasing the running application}

\begin{figure}[h!]
\centering
	{\includegraphics[width=0.8\textwidth]{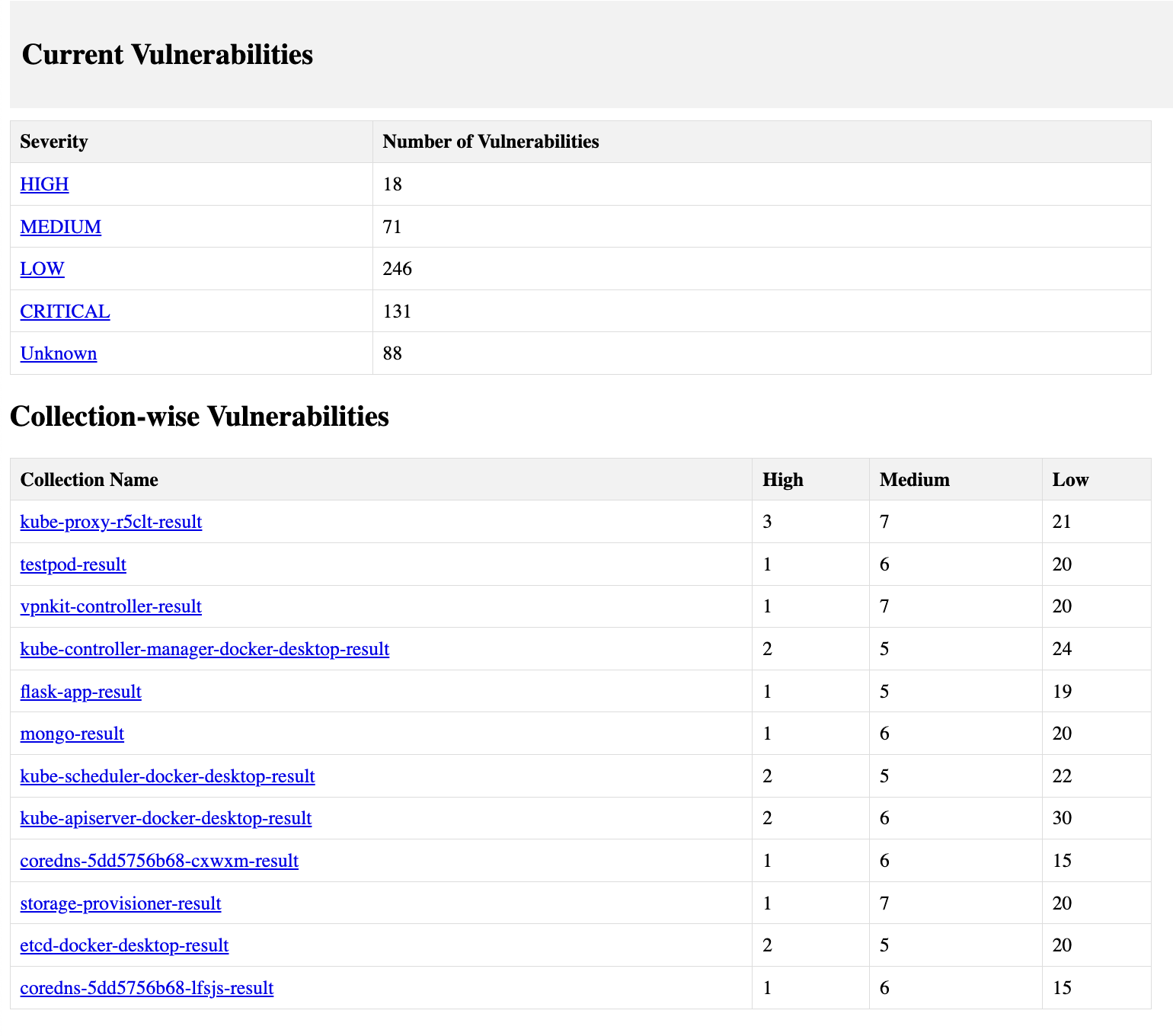}}
	\caption{Homepage of CLS}
\end{figure}

Figure 3 illustrates the homepage of the central logging system, providing a comprehensive overview of the cluster's current status. It highlights the total number of vulnerabilities, categorized by severity levels. Additionally, it details the existing vulnerabilities within each collection, including a breakdown of their respective severity counts.

\newpage

\begin{figure}[h!]
\centering
	{\includegraphics[width=\textwidth]{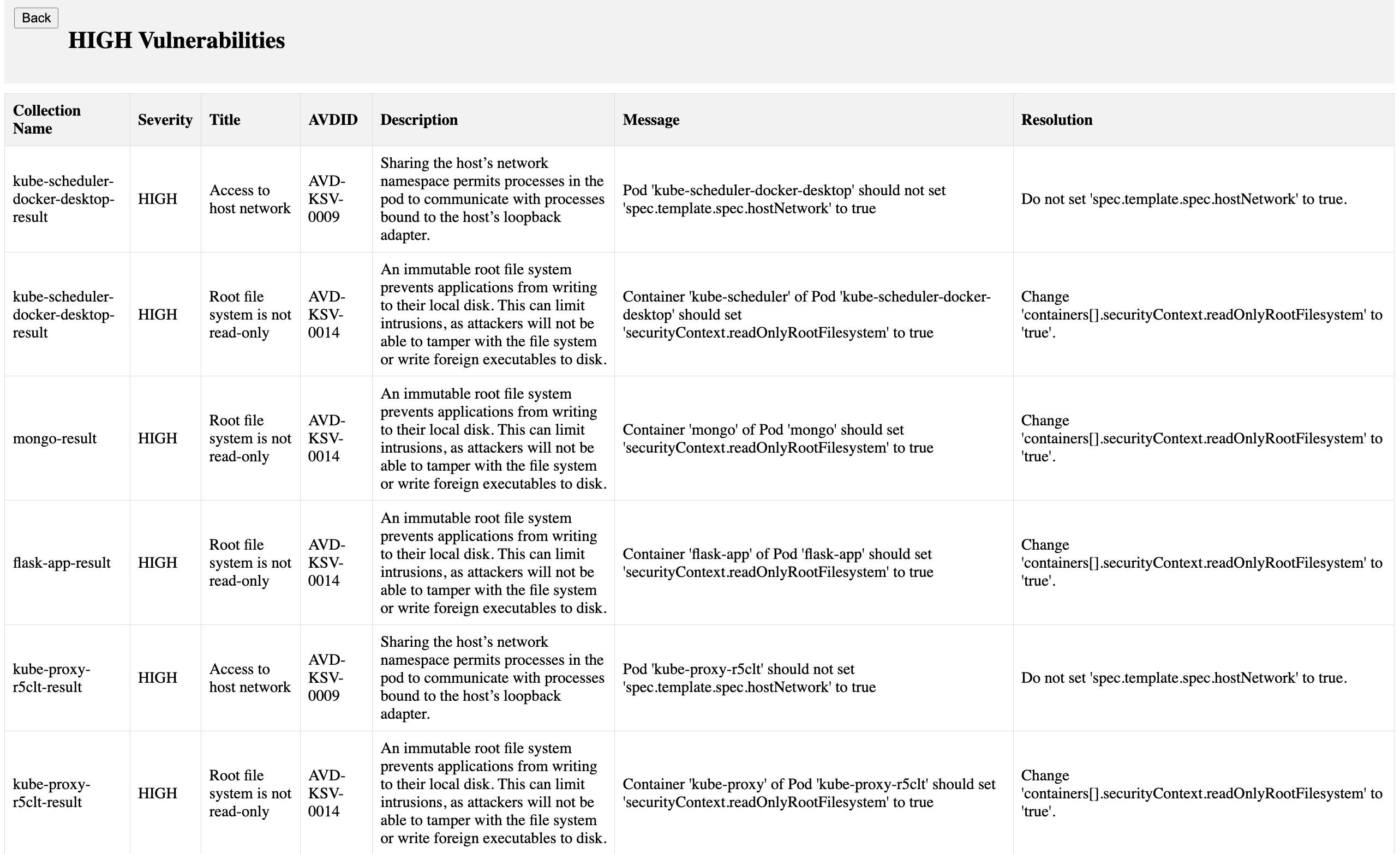}}
	\caption{Severity level page}
\end{figure}

Figure 4 presents the \textit{HIGH} severity page, which can be accessed by selecting \textbf{HIGH} from the homepage. This page displays all \textit{HIGH} misconfiguration vulnerabilities across the entire cluster. It provides detailed information for each vulnerability, including the name of the collection, its severity level, the title, the AVDID, a description, a message, and a recommended resolution that an administrator can implement to rectify the misconfiguration.

\newpage

\begin{figure}[h!]
\centering
	{\includegraphics[width=\textwidth]{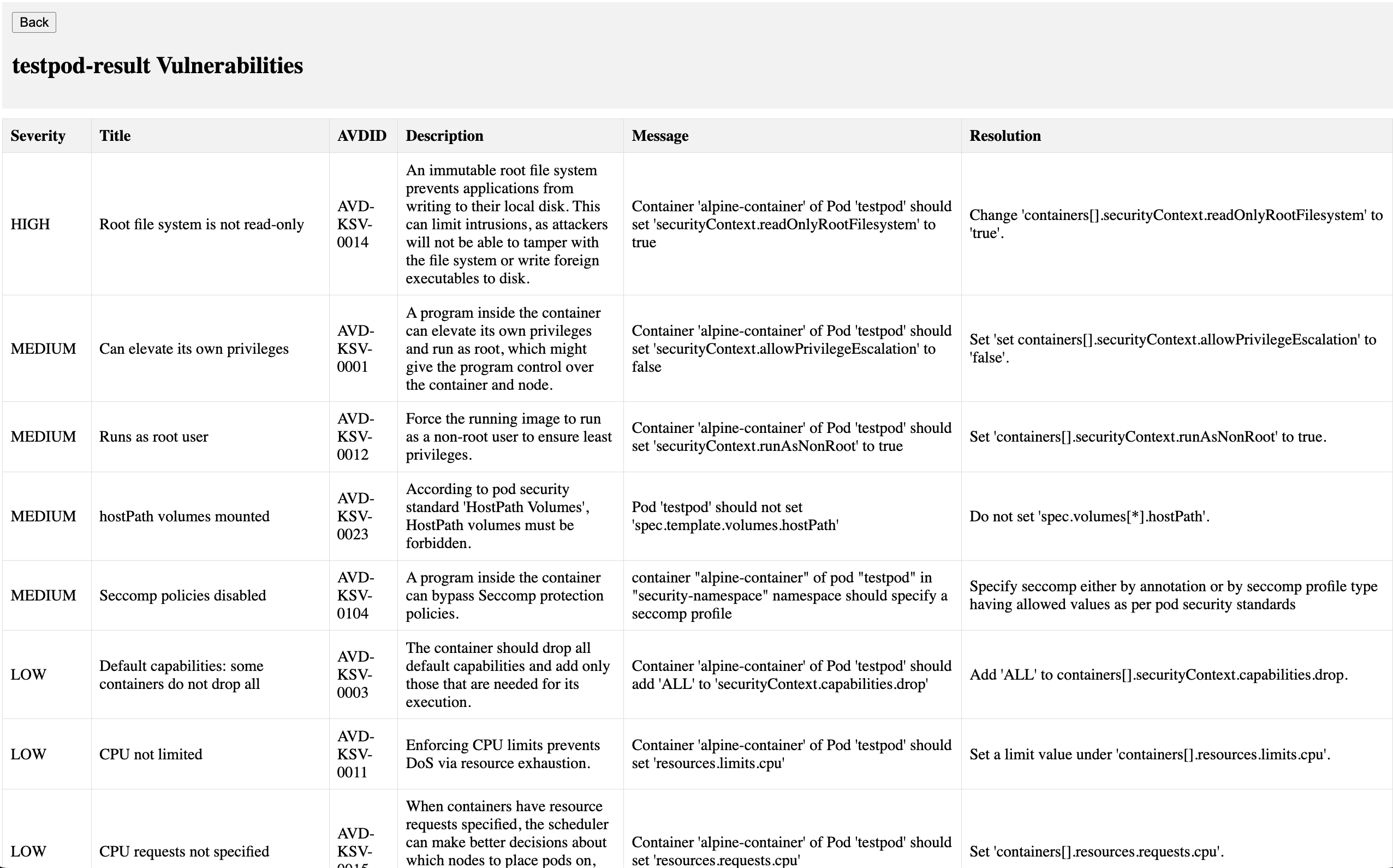}}
	\caption{Collection page}
\end{figure}

Figure 5 presents the collection page for the TestPod, providing a detailed view of the identified security misconfigurations for the single collection. This page is organized to display issues from highest to lowest severity and includes several key elements: the title, AVDID, a description of the misconfiguration, a detailed message regarding the issue, and the recommended resolution steps.

This section assesses the suggested model by using both quantitative metrics and qualitative analysis, concentrating on its effectiveness and performance. Furthermore, this section discusses whether the initial goals of this research were met and highlights the strengths and weaknesses of the proposed solution. It also details the challenges encountered during the paper.

\subsection{Examining the results}

The previous section outlines how the central logging solution is implemented, the outputs and metrics are then obtained and analysed in order to evaluate the reliability of the solution.

\begin{center}
\begin{table}[h!]
\small
\label{tab:execution_times}
\begin{tabularx}{\textwidth}{|C|C|C|C|C|C|}
\hline
\textbf{Script} & \textbf{Min Time (s)} & \textbf{Max Time (s)} & \textbf{Average Time (s)} & \textbf{Median Time (s)} & \textbf{Mode Time (s)} \\
\hline
run\_script.sh  & 29 & 35 & 29.95 & 30 & 30 \\
\hline
compare.py & 0 & 1 & 0.20 & 0 & 0 \\
\hline
convert\_jq.sh  & 0 & 1 & 0.75 & 1 & 1 \\
\hline
merged.sh & 0 & 1 & 0.30 & 0 & 0 \\
\hline
sender.py & 0 & 1 & 0.60 & 1 & 1 \\
\hline
\textbf{Total cycle time} & 30 & 37 & 32 & 32 & 32 \\
\hline
\end{tabularx}
\caption{Combined execution times detailing the minimum, maximum, average, median, and mode times for each script across 20 runs}
\end{table}
\end{center}

Table 2 summarizes the execution times of all tools and scripts over 20 runs, providing data on the minimum, maximum, average, median, and mode execution times. 
\begin{itemize}
    \item \textbf{\texttt{run\_script.sh}}: This represents the overall execution time for all tools combined, varying between 29 to 35 seconds. The longer duration is due to Trivy needing to download its database during the initial run. Subsequent runs, which do not require this download, typically see a reduction in execution time by 5-6 seconds.
    \item \textbf{\texttt{compare.py}}: This measures the time required to compare the outputs of the tools with each other, exhibiting an exceptionally low execution time of 0 to 1 seconds. It consistently demonstrates speed and efficiency.
    \item \textbf{\texttt{convert\_jq.sh}}: This indicates the time required to convert all outputs into a predefined template, which is remarkably fast, with execution times ranging from 0 to 1 seconds.
    \item \textbf{\texttt{merged.sh}}: This represents the time needed to merge all the output files from each of the tools into a single, larger file, ready for uploading to MongoDB.
    \item \textbf{\texttt{sender.py}}: This describes the duration required to upload the results to the MongoDB, with the execution times varying from 0 to 1 seconds.
\end{itemize}

The total cycle time spans from 30 to 37 seconds, with an average and most frequent execution time of 32 seconds, this reflects the combined impact of all the scripts on their overall processing, execution and completion time.

\subsection{Execution of each tool}

Each tool varies in execution time, so I aimed to assess the performance of each to determine which one takes the longest and if this should be considered a significant factor. One thing to note is that when running each tool individually, the pod manifest files must be gathered per tool, which impacts the time of each tools execution time. 

\begin{center}
\begin{table}[h!]
\small
\label{tab:trivy_times}
\begin{tabularx}{\textwidth}{|C|C|C|C|C|C|}
\hline
\textbf{Tool} & \textbf{Min Time (s)} & \textbf{Max Time (s)} & \textbf{Average Time (s)} & \textbf{Median Time (s)} & \textbf{Mode Time (s)} \\
\hline
Trivy & 24 & 29 & 24.3 & 24 & 24 \\
\hline
Kubescore & 4 & 5 & 4.8 & 5 & 5 \\
\hline
Kubesec & 6 & 8 & 6.9 & 7 & 7 \\
\hline
Kubelint & 5 & 6 & 5.8 & 6 & 6 \\
\hline
\end{tabularx}
\caption{Tool execution times detailing the minimum, maximum, average, median, and mode times across 20 runs}
\end{table}
\end{center}

Table 3 provides the execution times for the open source tools over 20 runs, illustrating their performance characteristics:

\textbf{Trivy}

Trivy consistently executed around 24 seconds, with its times ranging from a minimum of 24 to a maximum of 29 seconds, and an average of 24.3 seconds. The peak time of 29 seconds typically occurred during the initial run, as Trivy needed to download its database upon first execution, which prolonged the process. However, subsequent runs did not require a database download, resulting in quicker execution times.
\newpage
\textbf{Kubescore}

Kubescore had an execution time ranging from 4 to 5 seconds, with an average of 4.8 seconds. This makes it the lowest execution time of the open source tools tested in the central logging solution. 

\textbf{Kubesec}

Kubesec's execution times varied between 6 and 8 seconds, averaging 6.9 seconds. This made it quick and efficient at scanning manifest files. 

\textbf{Kubelint}

Kubelint's execution times spanned from 5 to 6 seconds, with an average of 5.8 seconds. This resulted in a rapid and efficient scan of manifest files.

\subsection{Security Tool Comparison}
This section provides a comparative analysis of the open-source security tools, focusing on their effectiveness in detecting misconfigurations and vulnerabilities. We will examine and contrast the outputs from each tool to understand their strengths and weaknesses in identifying potential security risks.

To ensure a balanced evaluation of the tools, I will analyze and review their outputs based on the "testpod" manifest file. This will involve presenting a snippet from each tool based on finding the same misconfiguration and then discussing their results.

\textbf{Trivy}

\begin{tcolorbox}[
    title=Trivy Sample Output,
    colback=white,
    colframe=black,
    sharp corners,
    width=\textwidth, 
    fonttitle=\bfseries,
    fontupper=\small, 
    breakable
]
\begin{verbatim}
{
  "Type": "Kubernetes Security Check",
  "ID": "KSV014",
  "AVDID": "AVD-KSV-0014",
  "Title": "Root file system is not read-only",
  "Description": "An immutable root file system prevents applications
                  from writing to their local disk. This can 
                  limit intrusions, as attackers will not be 
                  able to tamper with the file system or write 
                  foreign executables to disk.",
  "Message": "Container 'alpine-container' of Pod 'testpod' should set 
              'securityContext.readOnlyRootFilesystem' to true",
  "Resolution": "Change 
                'containers[].securityContext.readOnlyRootFilesystem' 
                to 'true'.",
  "Severity": "HIGH"
}
\end{verbatim}
\end{tcolorbox}

The Trivy scan of the "testpod" manifest file has uncovered a high-severity vulnerability (ID: KSV014, AVDID: AVD-KSV-0014) in the 'alpine-container' configuration within the pod. The container's root file system is writable, presenting a major security risk as it could allow unauthorized modifications and the installation of malicious executables. To address this issue, it is advised to set the securityContext.readOnlyRootFilesystem attribute of the container to true. This adjustment will enforce a read-only file system, significantly boosting the pod's security integrity. Trivy provides a clear and concise output, the severity rating helps administrators quickly identify and prioritise what is critical to securing the system. 
\newpage
\textbf{Kubescore}

\begin{tcolorbox}[
    title=Kubescore Sample Output,
    colback=white,
    colframe=black,
    sharp corners,
    width=\textwidth, 
    fonttitle=\bfseries,
    fontupper=\small, 
    breakable
]
\begin{verbatim}
  {
    "name": "Container Security Context ReadOnlyRootFilesystem",
    "id": "container-security-context-readonlyrootfilesystem",
    "comment": "Makes sure that all pods have a security 
               context with read only filesystem set",
    "grade": 1,
    "severity": "LOW"
  }
\end{verbatim}
\end{tcolorbox}

The KubeScore output checks if all pods have a read-only root file system set. It's assigned a severity of LOW and a grade of 1. The output is clear but could benefit from more detailed explanations for a better understanding. Notably, this LOW severity rating contrasts with the HIGH severity rating provided by Trivy for a similar security concern in the "testpod" manifest file. This difference highlights the need for administrators to critically assess and perhaps even seek multiple sources or tools when evaluating security issues, as each tool may have its own perspective on risk assessment. 

\textbf{Kubesec}

\begin{tcolorbox}[
    title=Kubesec Sample Output,
    colback=white,
    colframe=black,
    sharp corners,
    width=\textwidth, 
    fonttitle=\bfseries,
    fontupper=\small, 
    breakable
]
\begin{verbatim}
  {
    "id": "ReadOnlyRootFilesystem",
    "selector": 
    "containers[] .securityContext .readOnlyRootFilesystem == true",
    "reason": "An immutable root filesystem can prevent 
            malicious binaries being added to PATH and increase 
            attack cost",
    "points": 1,
    "severity": "LOW"
  }
\end{verbatim}
\end{tcolorbox}

The KubeSec output for "ReadOnlyRootFilesystem" checks if containers have the securityContext.readOnlyRootFilesystem set to true, aiming to secure containers by making the root file system immutable, rating it as LOW severity. In contrast, Trivy's more detailed output assigns a HIGH severity to a similar setting, providing broader context and implications. KubeScore, like KubeSec, also rates it LOW but without extensive details. KubeSec's concise style, similar to KubeScore's, focuses on direct security benefits but lacks the depth found in Trivy, which could be more useful for administrators making informed security decisions. This again highlights how one tool could influence how security measures are prioritised and implemented. 

\textbf{Kubelint}

\begin{tcolorbox}[
    title=Kubelint Sample Output,
    colback=white,
    colframe=black,
    sharp corners,
    width=\textwidth, 
    fonttitle=\bfseries,
    fontupper=\small, 
    breakable
]
\begin{verbatim}
  {
    "Check": "no-read-only-root-fs",
    "Message": "container \"alpine-container\" does not 
               have a read-only root file system",
    "Remediation": "Set readOnlyRootFilesystem to true 
                   in the container securityContext.",
    "Severity": "Unknown"
  }
\end{verbatim}
\end{tcolorbox}

The Kube-Linter output identifies that the "alpine-container" lacks a read-only root file system and recommends setting readOnlyRootFilesystem to true in the container's securityContext. Unlike Trivy, which assigns a HIGH severity to the same issue and provides extensive context, both KubeScore and KubeSec label this issue as LOW severity, indicating less urgency. However, Kube-Linter's "Unknown" severity could leave administrators unsure about how critically they need to address the misconfiguration identified. This highlights the discrepancies across different security tools, outlining the need for either a consistent security evaluation or else the use of multiple tools for a thorough security analysis.

Reviewing the outputs from Trivy, KubeScore, KubeSec, and Kube-Linter reveals significant differences in how each tool interprets and reports Kubernetes security vulnerabilities. While Trivy emphasizes the severity of issues by providing detailed context and a high urgency rating, KubeScore and KubeSec tend to offer more straightforward, less urgent assessments with a focus on practical remediation steps. Kube-Linter distinguishes itself by offering direct remediation advice but often leaves the severity as "Unknown," potentially complicating priority assessments for administrators. Each tool not only varies in its interpretation of security risks but also uses different keys in their JSON outputs, reflecting different approaches to structuring security data. 

\subsection{Challenges}

Comparing the outputs from the security tools: Trivy, KubeScore, KubeSec, and Kube-Linter had its challenges due to the varying keys used in their JSON outputs. Each tool organizes its data differently, using unique keys to denote aspects like severity, recommendations, and identification tags. This inconsistency in data structure complicated the central logging solution, as the data could not be filtered quickly. 

Additionally, each tool interprets and assigns different severity levels. Trivy assigned a HIGH severity to a vulnerability that KubeScore or KubeSec considered a LOW severity. Whereas Kube-Linter categorised it as an "Unknown" severity level, which made converting all the results to a predefined template more difficult. Also, if an administrator only used 1 security tool and it denoted a misconfiguration as LOW, then it could be ignored or skipped. 

\section{Conclusions and Future Work}

This paper presented a centralized logging solution aimed at mitigating Kubernetes misconfigurations using open-source tools. The proposed solution integrates multiple security tools to identify and log misconfigurations in Kubernetes manifests, providing a consolidated view for administrators. By automating the detection and aggregation of misconfiguration data, the solution significantly enhances the visibility of potential security issues within a Kubernetes cluster.

The implemented solution successfully achieved its primary goals: it demonstrated the feasibility of integrating various open-source tools into a unified logging system, evaluated the effectiveness of these tools in a real-world Kubernetes environment, and provided a detailed comparative analysis of their outputs. The central logging solution was able to identify critical security vulnerabilities, streamline the diagnostic process, and present the results in an accessible format via a web interface. This work contributes to the field of Kubernetes security by offering a practical tool for improving the security posture of Kubernetes deployments.

For an ideal central logging solution, the best scenario would involve a drag-and-drop feature that allows for easy integration of new tools. However, the current system struggles with this because of the variability in the key/value pairs used in their JSON outputs. A potential future approach could involve using machine learning to categorise and reconcile these differing key/value pairs.

Another area for future development could concentrate on refining how messages and descriptions from each tool are compared to eliminate redundant results. Currently, this is handled using Python's "SequenceMatcher" library, which identifies similar string sequences without considering the context of the messages. Implementing a machine learning strategy could improve the effectiveness of these comparisons by analyzing the context within the messages.

In today's technological landscape, containerised applications are increasingly popular, with Kubernetes being adopted by major organizations like Google, Adidas, Spotify, Bose, and the United States Department of Defense. This widespread adoption has led to the emergence of security vulnerabilities, particularly those stemming from cluster misconfigurations. This dissertation has concentrated on developing a centralized logging solution that leverages popular open-source tools to identify misconfigurations in Kubernetes manifest files. Prior to this paper, there was no unified solution that integrated multiple open-source tools. 

This solution serves as a proof of concept for those interested in delving deeper into misconfigurations detected by open-source tools in Kubernetes environments. There is potential for further enhancements to elevate this solution to a production-grade level, making it an invaluable resource for administrators to identify and address security vulnerabilities effectively.

\newpage
\bibliographystyle{elsarticle-num}
\bibliography{References.bib}

\end{document}